\documentclass[10pt,conference]{IEEEtran}

\usepackage{research5}
\usepackage{psfrag}
\usepackage{epsfig}
\usepackage{graphicx}
\usepackage{multicol}

\allowdisplaybreaks[3]

\newtheorem{thm}{Theorem}

\newcommand{\Ber}[1]{\textnormal{Ber}\left(#1\right)}

\title{The Poisson Channel with Side Information}

\author{\authorblockN{Shraga Bross}
\authorblockA{School of Enginerring\\
Bar-Ilan University, Israel\\
\texttt{ brosss@macs.biu.ac.il}}\and
\authorblockN{Amos Lapidoth~~~~~~~Ligong Wang}
\authorblockA{Signal and Information Processing Laboratory\\
ETH Zurich, Switzerland\\
\texttt{\{lapidoth, wang\}@isi.ee.ethz.ch}}}

\date{}

\begin{document}
\maketitle

\begin{abstract}
  The continuous-time, peak-limited, infinite-bandwidth Poisson
  channel  with spurious counts is considered. It is shown that if
  the times at which the spurious counts occur are known noncausally to 
  the transmitter but not to the receiver, then the capacity is
  equal to that of the 
  Poisson channel with no spurious counts. 
  Knowing the times at which the
  spurious counts occur only causally at the transmitter does not
  increase capacity. 
\end{abstract}

\section{Introduction}

Communication aided by channel-state information (CSI) at the
transmitter provides a rich source for interesting problems
in information theory. A discrete memoryless channel (DMC) with
independent and identically distributed (i.i.d.) states
is described by a transition law
\begin{subequations}\label{eq:DMC_S}
\begin{equation}
  \Pr (Y=y|X=x,S=s) = W(y|x,s)
\end{equation}
and  a state law
\begin{equation}
  P_S(s),
\end{equation}
\end{subequations}
where $X$ denotes the channel input, $Y$ denotes the channel output,
and $S$ denotes the state of the channel.

Shannon \cite{shannon58} studied the case where the CSI is unknown to
the receiver and known
\emph{causally} to the transmitter. In this scenario, before sending
$x_k$, the transmitter knows $\{s_1,\ldots,s_k\}$. 
He showed that 
the capacity of the channel \eqref{eq:DMC_S}  is
given by
\begin{equation}\label{eq:shannon}
  \sup_{P_U}I(U;Y),
\end{equation}
where $U$ is a random variable taking value in the set of mappings
from channel states to channel inputs, and where $P_{Y|U}$ is given by
\begin{equation*}
  P_{Y|U}(y|u) = \sum_s P_S(s) W(y|u(s),s).
\end{equation*}

A different scenario is where the CSI is still unknown to the receiver but
known \emph{noncausally} to the
transmitter. In this case, the transmitter knows the whole state
sequence before starting to transmit. The capacity of \eqref{eq:DMC_S} in this
case was found by Gel'fand and Pinsker \cite{gelfandpinsker80_3} to be
\begin{equation}\label{eq:gelfandpinsker}
  \sup_{U\markov (X,S) \markov Y} I(U;Y)-
  I(U;S),
\end{equation}
where the supremum is over all joint distributions of the form
\begin{equation*}
  P_U(u)P_{X,S|U}(x,s|u)W(y|x,s)
\end{equation*}
whose marginal distribution on $S$ is $P_S(\cdot)$.

The capacity with noncausal CSI was computed in
various cases. The case of writing on binary memory
with defects was solved by Kusnetsov and Tsybakov
\cite{kusnetsovtsybakov74} before the discovery of the
formula~\eqref{eq:gelfandpinsker}; the case of writing on binary
memory with defects \emph{and noise} was solved by Heegard and El
Gamal \cite{heegardelgamal83}; and the case of additive Gaussian noise
channel with additive states (``writing on dirty paper'')
was solved by Costa \cite{costa83}. In these cases, capacity was found
to be equal to the capacity when 
there are no states at all.

In the present work, we consider the continuous-time, peak-limited
Poisson channel with CSI at the transmitter. This channel model
without states was studied in
\cite{kabanov78}, \cite{davis80}, and \cite{wyner88}. In \cite{brossshamai03}, Bross and
Shamai considered the Poisson arbitrarily varying channel (AVC) with states.
In their set-up the states correspond to a nonnegative signal that
is added to the channel input.

We consider a model where the channel states correspond to spurious
counts at the receiver, and where the
receiver does not know which counts are spurious. We study first the case
where the CSI is known noncausally to the transmitter, and then
the case where the CSI is known causally to the transmitter.

In the noncausal
CSI case, we distinguish between two settings. In the first setting we
assume that the states are chosen by an adversary 
subject to a constraint on the average number of spurious counts per
second, and we allow the transmitter and the
receiver to use random codes. Since the state sequence can be
arbitrary, we cannot use 
Gel'fand and Pinsker's formula \eqref{eq:gelfandpinsker}. Instead,
as in \cite{kusnetsovtsybakov74} and \cite{heegardelgamal83}, we show that
the capacity with no spurious counts can be achieved on this channel
with random codes by construction. In the second setting we assume
that the
spurious counts are random (not 
necessarily a homogeneous Poisson process). Using the result from the
first setting,
we show that the capacity with no spurious counts is achievable on
this channel with deterministic codes.

For the causal CSI setting, we show that, even if the spurious counts
obey a homogeneous Poisson law, causal
CSI does not increase  
the capacity of this channel. Thus, as in \cite{kusnetsovtsybakov74}
and \cite{heegardelgamal83}, in our channel causal
CSI does not increase 
capacity at all, while noncausal CSI increases it to that of
the same channel model but without states.

The rest of this paper is arranged as follows: in Section
\ref{sec:normal} we recall some important results regarding the
peak-limited Poisson channel; in Section \ref{sec:nsi} we state and
prove the capacity results for the noncausal CSI case; and
in Section \ref{sec:causal} we discuss the causal CSI case.

\section{Preliminary: the Poisson Channel}\label{sec:normal}

In this section we recall some important results regarding the
peak-limited Poisson channel. This channel's time-$t$ 
input $x(t)$ is a nonnegative real, and its time-$t$ output $y(t)$ is a
nonnegative integer. For a given input signal $x(t)$, $t\in \Reals$,
the output $Y(t)$, $t\in \Reals$ is a Poisson process whose time-$t$
intensity is $x(t)+\lambda$, where $\lambda$ is a nonnegative constant
called the \emph{dark current}. 

We impose a peak-power constraint on the input so
\begin{equation}\label{eq:peak}
  x(t)\le A, \quad t\in\Reals,
\end{equation}
where $A>0$ is the maximal allowed input power. We denote the capacity
of this channel (in bits per second) by $C(A,\lambda)$. 

The value of $C(A,\lambda)$ was first found by Kabanov
\cite{kabanov78} and Davis \cite{davis80} using martingale
techniques. Later, Wyner \cite{wyner88} showed that
$C(A,\lambda)$ can be achieved by dividing the channel into small
time-slots and then looking at the resulting DMC.

To derive an achievability result, we follow Wyner and discretize this
channel in time with every 
time-slot being  
$\Delta$ seconds long. We further restrict the input distribution so that
within each time-slot the input is constant: either $A$ or
$0$. The receiver front-end produces $0$ if there were no counts in the
time-slot, and $1$ if there were one or more counts. This
discrete-time channel is 
memoryless and for small $\Delta$ its law can
be approximated by  
\begin{equation}\label{eq:discrete}
  W(1|x)=\begin{cases} \lambda\Delta, & x=0 \\ (A+\lambda)\Delta, & x=1. \end{cases}
\end{equation}
We denote the capacity of the channel~\eqref{eq:discrete} (in bits per
channel use) by $C_\Delta (A, \lambda)$. Wyner
\cite{wyner88} showed that
\begin{IEEEeqnarray*}{rCl} 
  C(A, \lambda) & = & \lim_{\Delta\downarrow 0}
  \frac{C_\Delta(A,\lambda)}{\Delta} \\
  & = & \max_{p\in(0,1)} \Big\{ p(A+\lambda)\log(A+\lambda) +
    (1-p)\lambda\log \lambda \\
    && ~~~~~~~~~~~~~~{}- (pA + \lambda)\log(pA+\lambda)\Big\}.
\end{IEEEeqnarray*}
We note that $C(A,\lambda)$ can also be written as
\begin{equation}\label{eq:normal_capacity}
  C(A,\lambda) = \max_{p\in(0,1)} (pA+\lambda) D\Bigg(
      \left.\Ber{\frac{p(A+\lambda)}{pA+\lambda}}\right\|  
        \Ber{p} \Bigg),
\end{equation}
where $\Ber{\pi}$ denotes the Bernoulli distribution of parameter~$\pi$.

\section{Noncausal CSI}\label{sec:nsi} 

We now consider the continuous-time Poisson channel as described in
Section~\ref{sec:normal}, but with spurious counts occurring at the
receiver. We assume that the times of these occurrences are known
noncausally to the 
transmitter but not to the receiver. We consider two settings: for
Section~\ref{sub:random} we 
assume that the spurious counts are generated by a malicious adversary
and that the transmitter and the receiver are allowed to use random
codes; for Section~\ref{sub:deter} we assume that the spurious
counts occur randomly according to a law known to both transmitter and
receiver and that only deterministic codes are allowed. 

\subsection{Random Codes against Arbitrary States}\label{sub:random}

Consider the Poisson channel as described in Section~\ref{sec:normal}
but with an adversary who generates spurious counts at the receiver and
reveals the times at which they occur to the transmitter before
communication begins. These spurious 
counts can be modeled by a counting signal $s(\cdot)$, which is a
nonnegative, integer-valued, monotonically increasing function.
Thus, conditional on the input being $x(\cdot)$, the
output is given by 
\begin{equation}\label{eq:model_s}
  Y(t) = Z(t)+s(t),\quad t\in\Reals,
\end{equation}
where $Z(t)$ is a Poisson process whose time-$t$ intensity is
$x(t)+\lambda$. We assume that
the adversary is subject to the restriction that, within each
transmission block, the average number of
spurious counts per second cannot exceed a certain constant $\nu$ which
is known to both transmitter and receiver.

In a $(T, R)$ (deterministic) code, the encoder
maps the message $M\in\{1,\ldots,2^{RT}\}$ and the channel state
$s(t)$, $t\in[0,T]$, to an input signal $x(t)$, $t\in [0,T]$, and
the decoder guesses the message $M$ from the channel output
$y(t)$, $t\in [0,T]$. A $(T, R)$ \emph{random code} is a
probability distribution on all deterministic $(T, R)$ codes.\footnote{For
more explicit formulations of random and deterministic codes, see
\cite{lapidothnarayan98} and references therein.}

A rate $R$ (in bits per second) is said to be \emph{achievable with
  random codes} on the channel \eqref{eq:model_s} if, for every $T>0$, 
there exists a random $(T, R)$ code such that, as $T$ tends to
infinity, the average probability of a guessing error tends to zero
for \emph{all possible} $s(\cdot)$. 
The \emph{random coding
  capacity} of this 
channel is defined as the supremum over all rates achievable with
random codes.

Since the adversary can choose not to introduce any spurious counts,
the random coding capacity 
of the channel \eqref{eq:model_s} is upper-bounded by $C(A,\lambda)$,
which is given in 
\eqref{eq:normal_capacity}. 
Our first result is that this bound is tight:

\begin{thm}\label{thm:muT}
  For any positive $A$, $\lambda$ and $\nu$, the random coding
  capacity of the channel \eqref{eq:model_s}, where $s(\cdot)$
  is known noncausally to the transmitter but unknown to the
  receiver, is equal to $C(A,\lambda)$. 
\end{thm}
\begin{proof}
  We only need to prove the lower bound.
  Namely, we need to show that any rate below the right-hand side (RHS)
  of~\eqref{eq:normal_capacity} is 
  achievable with random codes on the channel \eqref{eq:model_s}. To
  this end, for fixed positive constants $T$, $R$ and $\alpha$, we
  shall construct a block
  coding scheme to transmit $RT$ bits of information using
  $(1+\alpha)T$ seconds. (Later we shall choose $\alpha$ arbitrarily
  small.) We divide the block into
  two phases, first the \emph{training phase} and then the \emph{information
    phase}, where the training phase is $\alpha T$ seconds long and
  the information phase is $T$ seconds long. Within each phase we
  make the  
  same discretization in time as in  
  Section~\ref{sec:normal}, where every time-slot is $\Delta$ seconds
  long. We choose $\Delta$ to be small enough so that $\lambda\Delta$,
  $A\Delta$ and $\nu\Delta$ 
  are all small compared to one. In this case, our channel is reduced
  to a DMC whose transition law can
  be approximated by: 
\begin{equation}\label{eq:discrete_muT}
  W(1|x,s) = \begin{cases} \lambda\Delta, & x=0, s=0 \\ (A+\lambda)\Delta, & x=1, s=0 \\ 1, & s=1. \end{cases}
\end{equation}
Here $x=1$ means that the continuous-time channel input in the
time-slot is $A$, and $x=0$ means that the continuous-time channel
input is zero; $y=1$ means that at least one count is
detected at the receiver in the 
time-slot, and $y=0$ means that no counts are detected; $s=1$ means that
there is at least one spurious count in the time-slot, so 
the output is stuck at $1$, and $s=0$ means that there are no spurious counts
in the time-slot, so the channel is the same
as~\eqref{eq:discrete}.  From now on we shall refer to time-slots
where $s=1$ as ``stuck slots.''

Denote the total number of stuck slots in the information phase by
$\mu T$. Then, by the state constraint, 
\begin{equation}\label{eq:munu}
  \mu\le (1+\alpha)\nu.
\end{equation}
In the
training phase the transmitter tells the receiver the value of $\mu
T$. To do this, the transmitter and the receiver use the channel as an
AVC. Namely, in this phase the
transmitter ignores his 
knowledge about the times of the spurious counts. Since, by the state
constraint, the total number of stuck slots in the whole transmission
block cannot exceed $(1+\alpha)\nu T$, we know that the total number
of stuck slots in the training phase also cannot exceed $(1+\alpha)\nu
T$. It can be easily 
verified using the formula for random coding capacity of the AVC with 
state constraints \cite{csiszarnarayan88} that the random coding
capacity of the AVC \eqref{eq:discrete_muT} under this constraint is
proportional to $\Delta$ for small $\Delta$. Thus, the amount of
information that can be reliably transmitted in the training phase is
proportional to $\alpha T$ for large $T$ and small $\Delta$. On the
other hand, according to \eqref{eq:munu}, we only need $\log\bigl(
(1+\alpha)\nu T\bigr)$ bits to describe $\mu T$. Thus we conclude that, for
any $\alpha>0$, for large enough $T$ and small enough $\Delta$, the
training phase can be accomplished successfully with high probability.

We next describe the information phase (which is $T$ seconds long). If
the training phase is 
successful, then in the information phase,
both the transmitter and the receiver know the
total number of stuck slots $\mu T$, but only the transmitter
knows their positions. With such knowledge, they can use the following
random coding scheme to transmit $RT$ bits of information in this phase:
\begin{itemize}
  \item \textbf{Codebook:} Generate $2^{(R+R')T}$ codewords
    independently such that the symbols within every codeword are
    chosen i.i.d. $\Ber{p}$. Label the codewords as
    $\vect{x}(m,k)$ where $m\in \{1,\ldots, 
    2^{RT}\}$ and $k\in\{ 1,\ldots, 2^{R'T}\}$. 
  \item \textbf{Encoder:} For a given message $m \in \{1,\ldots,
    2^{RT}\}$ and a state sequence $\vect{s}$, find a $k$ such that
    \begin{equation*}
      \sum_{j=1}^{T/\Delta}
      \textnormal{I}\Big\{\bigl(x_j(m,k),s_j\bigr)=(1,1)\Big\}  \ge 
      (1-\epsilon)\frac{p(A+\lambda)} { pA+\lambda} \mu T,
    \end{equation*}
    where $\textnormal{I}\{\cdot\}$ denotes the indicator function so the
    left-hand side (LHS) is the number of slots where
    $\vect{x}(m,k)$ and $\vect{s}$ 
    are both one.
    Send $\vect{x}(m,k)$. If no such
    $k$ can be found, send an arbitrary sequence. 
  \item \textbf{Decoder:} Find a codeword $\vect{x}(m', k')$ in the codebook
    such that, for the observed $\vect{y}$,
    \begin{IEEEeqnarray*}{rCl}
      \lefteqn{\sum_{j=1}^{T/\Delta}
      \textnormal{I}\Big\{\bigl(x_j(m',k'),y_j\bigr)=(1,1)\Big\}} \\
    &&~~~~~~~~~~~~~~\ge
      (1-\epsilon)\frac{p(A+\lambda)} { pA+\lambda} (pA+\lambda+\mu) T.
    \end{IEEEeqnarray*}
    Output
    $m'$. If no such codeword can be found, declare an error. 
\end{itemize} 
We next analyze the error probability of this random codebook. There are
three types of errors which we analyze separately:
\begin{itemize}
  \item The encoder cannot find a $k$ such that $\vect{x}(m,k)$ meets the
    requirement. We know that the total
    number of stuck slots in this phase is $\mu T$. Since
    the codebook is generated independently of the stuck 
    slots, we know that the symbols of a particular
    codeword 
    at these slots are drawn i.i.d. according to
    $\Ber{p}$. By Sanov's theorem
    \cite{sanov57,coverthomas91}, for large $T$, the 
    probability that a particular codeword satisfies the requirement,
    i.e., has at least $(1-\epsilon)\frac{p(A+\lambda)}{pA +
      \lambda}\mu T$ ones in these $\mu T$ slots, is
    approximately 
    $2^{\mu T
      D\bigl( \left.\Ber{(1-\epsilon)\frac{p(A+\lambda)}{pA+\lambda}}\right\| 
        \Ber{p}\bigr)}$. Therefore, when $T$ tends
    to infinity, the probability of this error tends to zero if we
    choose
    \begin{equation}\label{eq:R'}
      R' > \mu
      D\Bigg(\left.\Ber{(1-\epsilon)\frac{p(A+\lambda)}{pA+\lambda}}\right\| 
        \Ber{p}\Bigg). 
    \end{equation}
  \item There are less than $(1-\epsilon)\frac{p(A+\lambda)}{pA +
      \lambda}(pA+\lambda+\mu) T$ slots 
    where the transmitted codeword and the output are both equal to
    one. The probability of this error tends to zero as $T$ tends to
    infinity by the law of large numbers.
  \item There is some $\vect{x}'$ which is not the transmitted
    codeword such that there are at least
    $(1-\epsilon)\frac{p(A+\lambda)}{pA + \lambda}(pA+\lambda+\mu) T$
    slots where $x_j'=1$ and $y_j=1$. To analyze the probability of
    this error, 
    we first note that, by the law of large numbers, when $T$ is
    large, the number of slots 
    where $y_j=1$ is close to $(pA +\lambda + \mu)T$. We also note that a
    particular codeword that 
    is not transmitted is drawn i.i.d. $\Ber{p}$ independently of
    $\vect{y}$. Therefore, again by
    Sanov's theorem, the probability that this codeword has at
    least $(1-\epsilon)\frac{p(A+\lambda)}{pA +
      \lambda}(pA+\lambda+\mu) T$ ones at the approximately $(pA
    +\lambda + \mu)T$ slots where $y_j=1$ 
    is approximately $2^{(pA + \lambda + \mu) T
      D\bigl( \left.\Ber{(1-\epsilon)\frac{p(A+\lambda)}{pA+\lambda}}\right\| 
        \Ber{p}\bigr)}$. Thus, when $T$ tends to infinity, this
      probability tends to zero if we choose 
    \begin{IEEEeqnarray}{rCl}
      \lefteqn{R+R'<(pA+\lambda+\mu) \cdot}\nonumber\\
      &&~~~~~~D\Bigg(
      \left.\Ber{(1-\epsilon)\frac{p(A+\lambda)}{pA+\lambda}}\right\|  
        \Ber{p} \Bigg).\IEEEeqnarraynumspace\label{eq:RR'}
    \end{IEEEeqnarray}
\end{itemize}
By combining~\eqref{eq:R'} and~\eqref{eq:RR'} and noting that
$\epsilon$ can be chosen to be arbitrarily small, we conclude that,
for every $p\in(0,1)$, when $T$ is large and when $\Delta$ is small,
successful transmission in the information phase can be achieved with
high probability as long as
\begin{equation}\label{eq:rate}
  R<(pA+\lambda) D\Bigg(
      \left.\Ber{\frac{p(A+\lambda)}{pA+\lambda}}\right\|  
        \Ber{p} \Bigg).
\end{equation}
Furthermore, since we have shown that the training phase can be
accomplished with any positive $\alpha$, the overall
transmission rate, which is equal to
$\frac{R}{1+\alpha}$, can also be made arbitrarily close to the RHS of
\eqref{eq:rate}. Optimization over $p$ implies
that we can achieve all rates up to the RHS of
\eqref{eq:normal_capacity} with random coding. 
\end{proof}

\subsection{Deterministic Codes against Random States}\label{sub:deter}

We next consider the case where, rather than being an arbitrary
sequence chosen by an adversary, the spurious counts are random. Such
random 
counts can be modeled by a random counting process $S(\cdot)$ which
is independent of the message, so the
channel output is given by
\begin{equation}\label{eq:model_random}
  Y(t) = Z(t) + S(t),\quad t\in\Reals,
\end{equation}
where $Z(\cdot)$ is a Poisson process whose time-$t$ intensity is
$x(t)+\lambda$, and conditional on $x(\cdot)$, $Z(\cdot)$ is
independent of $S(\cdot)$. We assume that $S(0)=0$ with probability
one and 
\begin{equation}
  \varlimsup_{t\to\infty}\frac{\E{S(t)}}{t}<\infty. \label{eq:14}
\end{equation}
Note that these conditions are satisfied,
for example, when $S(\cdot)$ is a homogeneous Poisson process. We also
assume that the law of 
$S(\cdot)$ 
is known to both transmitter and receiver (and, in particular, the code
may depend on the law of $S(\cdot)$), while the realization of
$S(\cdot)$ is known noncausally to the transmitter but unknown to
the receiver. A rate is said to be
achievable on this channel if, for every $T>0$, there exists a
deterministic $(T,R)$
code such that the average probability of error averaged over
$S(\cdot)$ tends to zero as $T$ tends to infinity. The
capacity of this channel is defined as the supremum over all
achievable rates.

\begin{thm}
  The capacity of the channel \eqref{eq:model_random}, where the
  realization of $S(\cdot)$ is known noncausally to the transmitter
  but unknown to the receiver, is equal to $C(A,\lambda)$,
  irrespective of the law of $S(\cdot)$.
\end{thm}
\begin{proof}
  We first observe that the capacity of
  \eqref{eq:model_random} cannot exceed $C(A,\lambda)$. This is
  because we can mimic the channel \eqref{eq:model_random} over a
  channel without spurious counts as follows: The transmitter and
  the receiver use common randomness (which does not help on the
  single user channel without states) to generate $S(\cdot)$ and then
  the receiver 
  ignores its realization. 

  We shall next show that any
  rate below $C(A,\lambda)$ is achievable on \eqref{eq:model_random}. 
  Fix any $\epsilon>0$ and $R<C(A,\lambda)$. Let
  \begin{equation*}
    \zeta \triangleq \varlimsup_{t\to\infty}\frac{\E{S(t)}}{t}.
  \end{equation*}
  Since $\zeta<\infty$, there exists a $t_0$ such that
  \begin{equation*}
    \E{S(t)} \le 2 \zeta,\quad t>t_0.
  \end{equation*}
  Using this and Markov's inequality
  we have
  \begin{equation*}
    \Pr\left[S(t)\ge \frac{2\zeta}{\epsilon}\right] \le \epsilon,
    \quad t>t_0.
  \end{equation*}
  Thus the error probability of a $(T,R)$ random code where $T>t_0$
  can be bounded as
  \begin{IEEEeqnarray*}{rCl}
    \Pr[\textnormal{error}] & \le & \Pr\left[\textnormal{error},S(T)\ge
      \frac{2\zeta}{\epsilon}\right] + \Pr\left[\textnormal{error},
      S(T)< \frac{2\zeta}{\epsilon}\right]\\
    & \le & \Pr\left[S(T)\ge \frac{2\zeta}{\epsilon}\right] \\
    && {} +
    \Pr\left[\textnormal{error}\left| 
      S(T)< \frac{2\zeta}{\epsilon}\right.\right] \cdot \Pr
  \left[S(T)< \frac{2\zeta}{\epsilon}\right] \\
    & \le & \epsilon + \Pr\left[\textnormal{error}\left|
      S(T)< \frac{2\zeta}{\epsilon}\right.\right],\quad
  T>t_0.\IEEEyesnumber\label{eq:1} 
  \end{IEEEeqnarray*}
  To bound the second term on the RHS of \eqref{eq:1} we use
  Theorem~\ref{thm:muT}  
  which says that there exists $t_1$ such that, for
  any $T>t_1$, there exists a random $(T,R)$ code whose
  average error probability conditional on any realization of
  $S(\cdot)$ satisfying $s(T)<\frac{2\zeta}{\epsilon}$ is not larger
  than $\epsilon$. Therefore, for such codes,
  \begin{equation}\label{eq:2}
    \Pr\left[\textnormal{error}\left|
      S(T)< \frac{2\zeta}{\epsilon}\right.\right] \le \epsilon,\quad
  T>t_1.
  \end{equation}
  Combining \eqref{eq:1} and \eqref{eq:2} yields that for any
  $T>\max\{t_0,t_1\}$ there exists a random $(T,R)$ code 
  for which
  \begin{equation*}
    \Pr[\textnormal{error}] \le 2\epsilon.
  \end{equation*}
  Since this is true for all $\epsilon>0$ and $R<C(A,\lambda)$, we
  conclude that all rates below $C(A,\lambda)$ are achievable on
  \eqref{eq:model_random} with
  random codes.

  We next observe that we do not need to use random codes. Indeed,
  picking for every $T$ and $R$ the best deterministic $(T,R)$ code
  yields at worst 
  the same average error probability as that of any random $(T,R)$ code. Thus
  we conclude that
  any rate below $C(A,\lambda)$ is achievable on
  \eqref{eq:model_random} with deterministic codes, and hence the
  capacity of \eqref{eq:model_random} is equal to $C(A,\lambda)$.
\end{proof}

\section{Causal CSI}\label{sec:causal}

In this section we shall argue that causal CSI
does not increase the capacity of a peak-limited Poisson
channel. We look at the simplest case where the spurious
counts occur as a homogeneous Poisson process of intensity $\mu$. We shall
be imprecise regarding the definition of causality 
in continuous time by directly looking at the
DMC~\eqref{eq:discrete_muT}.\footnote{For formulations of causality in
  continuous time see \cite{lapidoth93} or \cite{kadotazakaiziv71}.}
Since the continuous-time state $S(\cdot)$ is a Poisson process, the
discrete-time state sequence $\vect{S}$ is i.i.d. with each component
taking the value 
$1$ with probability $\mu\Delta$ and taking the value $0$ otherwise.


To argue that having causal CSI does not increase capacity, we shall
show that every mapping $u$ from channel states to 
input symbols (as in \eqref{eq:shannon}) is equivalent to  
a \emph{deterministic input symbol} in the sense that it induces the
same output distribution as the latter. Indeed, since when $s=1$ the
input symbol $x$ has no influence on the output $Y$, 
we know that the value of $u(1)$ does not influence the output
distribution. Therefore
$u\colon s\mapsto u(s)$ is equivalent to the mapping that maps both $0$ and
$1$ to $u(0)$, and is thus also equivalent to the
deterministic input symbol $x=u(0)$.

In a more explicit argument, we use the capacity expression
\eqref{eq:shannon}. For 
any distribution $P_U$ on 
$\set{U}$, we let 
\begin{equation*}
  P_X(x) = \sum_{u\colon u(0)=x} P_U(u).
\end{equation*}
Then, for the above $P_U$ and $P_X$, 
\begin{equation*}
  I(U;Y) = I(X;Y).
\end{equation*}
Therefore we have
\begin{equation*}
  \sup_{P_U} I(U;Y) \le \sup_{P_X} I(X;Y),
\end{equation*}
where the LHS is the capacity of the channel with causal CSI, and
where the RHS is the capacity of the channel with no CSI. Thus we
conclude that the capacity of our channel model with causal CSI is not
larger than that of the channel with no CSI.

\bibliographystyle{IEEEtran}           
\bibliography{/Volumes/Data/wang/Library/texmf/tex/bibtex/header_short,/Volumes/Data/wang/Library/texmf/tex/bibtex/bibliofile}

\end{document}